\documentclass{ifacconf}

\usepackage{graphicx}      
\usepackage{natbib}        

\usepackage{amsmath} 
\usepackage{amssymb}  
\usepackage{siunitx}
\usepackage{tikz}
\usepackage{filecontents}
\usepackage{balance}
\usepackage[eulergreek]{sansmath}
\usetikzlibrary{arrows,positioning,shapes,intersections,patterns}
\usepackage{pgfplots}
\usepgfplotslibrary{patchplots,fillbetween}
\pgfplotsset{colormap={myblackwhite}{[1pt]
             rgb255(0pt)=(0, 150, 0);
             rgb255(500pt)=(0, 255, 0);
             rgb255(501pt)=(255, 0, 0);
             rgb255(1000pt)=(255, 100, 0)},
			 point meta min=-1, %
			 point meta max=1, 
			 compat = 1.13,
			 ticklabel style = {font=\sansmath\sffamily\small},
			 legend style = {font=\sf\small},legend cell align=left,
			 title style={yshift=-9pt, font = \sf\normalsize},
			 ylabel style={font = \sf\small},
			 xlabel style={font = \sf\small},
			 width=\columnwidth,
			 height=6cm,
			 }		 
\theoremstyle{break}
\newtheorem{mydef}{Definition}
\newtheorem{mylemma}{Lemma}
\newtheorem{myremark}{Remark}
\newtheorem{prope}{Property}

\newcommand{\R}{\mathbb{R}}
\newcommand{\N}{\mathbb{N}}
\newcommand{\C}{\mathcal{C}}
\newcommand{\X}{\mathcal{X}}

\newcommand{\bm}[1]{{\boldsymbol{#1}}}
\DeclareMathOperator{\diag}{diag}
\DeclareMathOperator{\var}{var}
\DeclareMathOperator{\mean}{\mu}
\DeclareMathOperator{\expval}{E}

\DeclareMathOperator{\tr}{Tr}

\newcommand{\GP}{\mathcal{GP}}
\newcommand{\GN}{\mathcal{N}}
\newcommand{\g}{\bm g}
\newcommand{\ddqd}{\ddot{\bm q}_\bm{d}}
\newcommand{\dqd}{\dot{\bm q}_\bm{d}}
\newcommand{\qd}{\bm{q}_\bm{d}}
\newcommand{\ddqe}{\ddot{{\bm e}}}
\newcommand{\dqe}{\dot{{\bm e}}}
\newcommand{\qe}{{\bm e}}
\newcommand{\ddq}{\ddot{\bm q}}
\newcommand{\dq}{\dot{\bm q}}
\newcommand{\q}{\bm q}
\newcommand{\qc}{\bm{q_c}}

\newcommand{\x}{\bm x}

\usepackage{mathtools}
\mathtoolsset{showonlyrefs}

\begin{document}
\begin{frontmatter}

\title{Stable Model-based Control with Gaussian Process Regression for Robot Manipulators
} 

\thanks[footnoteinfo]{The research leading to these results has received funding from the European Research Council under the European Union Seventh Framework Program (FP7/2007-2013) / ERC Starting Grant ``Control based on Human Models (con-humo)'' agreement n\textsuperscript{o}337654.}

\author[First]{Thomas Beckers} 
\author[First]{Jonas Umlauft} 
\author[First]{Sandra Hirche}

\address[First]{Chair of Information-oriented Control (ITR), 
                        Department of Electrical and Computer Engineering,
                        Technical University of Munich, Germany 
                        (e-mail: \{t.beckers,jonas.umlauft,hirche\}@tum.de)}

\begin{abstract}                
Computed-torque control requires a very precise dynamical model of the 
robot for compensating the manipulator dynamics. This allows reduction of the controller's feedback gains resulting in disturbance attenuation and 
other advantages. Finding precise models for manipulators is often difficult with 
parametric approaches, e.g. in the presence of complex friction or flexible links. 
Therefore, we propose a novel computed-torque control law which consists 
of a PD feedback and a dynamic feed forward compensation part with Gaussian 
Processes. For this purpose, the nonparametric Gaussian Process regression infers
the difference between an estimated and the true dynamics. In contrast 
to other approaches, we can guarantee that the tracking error is stochastically 
 bounded. Furthermore, if the 
number of training points tends to infinity, the tracking error is 
asymptotically stable in the large. In simulation and with an experiment, 
we demonstrate the applicability of the proposed control law and that it 
outperforms classical computed-torque approaches in terms of tracking precision.
\end{abstract}


\end{frontmatter}

\section{Introduction}
In the last decades, various robot control schemes have been proposed and 
most of them can be considered as a subset of computed-torque control 
laws. Computed-torque, a special case of feedback linearization, 
transforms the nonlinear system into an equivalent linear system through a 
change of variables and a suitable control input. Computed-torque control is 
able to derive very effective robot controllers that appear in 
robust, adaptive and learning control schemes~\citep{siciliano2010robotics}. 
With 
an exact model of the manipulator, this control law can compensate the robot 
dynamics to achieve a low gain feedback term which is beneficial in many ways: 
it allows safe physical human-robot interaction, reduces energy consumption, 
avoids disturbance attenuation in presence of noise and avoids the saturation of 
the actuators \citep{nguyen2008learning}. Since the accuracy of the 
compensation 
depends on the precision of the model, the model building process is the key to achieve good performance.

The uncertainties of a model can be separated in structural and parametric 
variations. The structural uncertainties come from the lack of knowledge of the 
underlying true physics. The parametric uncertainties exist since the exact 
values of length, masses, etc. are often unknown. The classical approach is to derive a dynamic model from first order physic 
laws, e.g., with a CAD model of the manipulator and increase the feedback part 
of the control law to compensate for structural and parametric uncertainties 
until a desired performance is achieved~\citep{spong2008robot}. But the 
increased gains are undesirable (as explained above) and therefore deriving 
more accurate models is of high importance.

Classical system identification algorithms for computed 
torque take advantage of the fact that the inertia parameters are linear with 
respect to the inverse dynamics \citep{slotine1987adaptive}. Parametric models 
which cover all dynamics are hard to obtain, especially in the presents of 
friction, flexible links or environment interaction. Therefore, nonparametric 
learning approaches provide promising results \citep{deisenroth2015gaussian}.
Gaussian Process regression (GPR) is a supervised learning technique with 
several advantages: it requires only a minimum of prior knowledge for 
arbitrary complex function, generalizes well even for little training data 
and has a precise trade-off between data fitting and 
smoothing \citep{rasmussen2006gaussian}. A Gaussian Process (GP) connects every point of a continuous input space with a 
normally distributed random variable. Any finite group of those infinitely many 
random variables follows a multivariate Gaussian distribution. Based on this, 
the result is a powerful tool for nonlinear function regression without the need 
of much prior knowledge. In contrast to many other regression techniques, e.g. neural networks, 
GP modeling provides not only a mean function but also a measure 
for the model fidelity based on the distance to training data. The output is a
Gaussian distributed variable which is fully described by its mean and variance.

\citet{alberto2014computed} present a computed-torque controller 
with Gaussian Process regression where the stiffness of the system depends on the variance of the 
learned model. The inherent 
learning of variable loads of the manipulator is done 
by~\citet{williams2009multi}. \citet{nguyen2010using} present 
a hybrid learning approach which incorporates model knowledge. The mentioned works show empirically promising results for GP-based control of robotic manipulators. However, they neither 
guarantee stability of the closed loop nor do they examine the influence of the training 
points mathematically. 

The contribution of this paper is a novel computed torque control law for 
robotic manipulators using Gaussian Process regression, which guarantees stability and consistency of 
the regression. For this purpose, a GP learns the difference between an 
estimated model and the true robot from training trajectories. Afterwards, the 
control law uses GPR to compensate the unknown robot dynamics. The proposed 
method also abstains from feeding back joint accelerations (in comparison to, 
e.g.,~\citet{alberto2014computed}) as these are difficult to measure directly 
and often inject noise. The derived method guarantees that the tracking error 
is stochastically bounded around zero independent of the number of training 
data. If the number of training points tends to infinity, this bound becomes 
arbitrary small and the tracking error is asymptotically stable in the 
large.

The paper is structured as follows: Section~\ref{sec:Def}
introduces the model of a robotic manipulator and GPR. 
Section~\ref{sec:ctrl_law} proposes the control law, proofs it stability and 
explains training of GPs. An empirical evaluation based on simulation and 
experiment follows in Section~\ref{sec:SimandExp}.

\section{Background}
\label{sec:Def}
\subsection{Gaussian Process Regression}
Let\footnote{\textbf{Notation}:
Vectors and vector-valued functions are denoted with bold characters. Capital 
letters describe matrices. The expression~$\mathcal{N}(\mu,\Sigma)$ denotes the 
normal distribution with mean~$\mu$ and covariance~$\Sigma$. The Euclidean norm 
is given by~$\Vert\cdot\Vert$. The mean and variance of a random variable 
is written as~$\mean(\cdot)$ or~$\expval(\cdot)$ and~$\var(\cdot)$, respectively.
The minimum singular value of a matrix is denoted with~$\sigma_{min}(\cdot)$.~$I_n$ 
denotes the~$n\times n$ identity matrix.}
 $(\Omega, \mathcal{F},P)$ be a probability space with the sample space 
$\Omega$, the corresponding~$\sigma$-algebra~$\mathcal{F}$ and the probability 
measure~$P$. The set~$\X \subseteq \R^d$ with~$d\in\N^*$ denotes the index set. 
A Gaussian Process is a discrete or real valued function~$f_{GP}(\x, \omega)$ which 
is a measurable function of~$\omega\in\Omega$ with~$\x\in\X$. The process is fully described by a mean 
function~$m(\x)$ and a covariance function~$k(\x,\x^\prime)$ since 
it is Gaussian distributed for any fixed~$\x$ . The GP is denoted by
\begin{align}
&f_{GP}(\x) \sim \GP(m(\x),k(\x,\x^\prime)),\qquad \x,\x^\prime\in\X,\\
&m(\x)\colon\X\to\R,\,k(\x,\x^\prime)\colon\X\times \X\to\R.
\end{align}
The mean function is usually set to zero as no prior knowledge about the 
function is given. The covariance function is a measure for the interference of 
two states~$(\x,\x^\prime)$. 
Probably the most widely 
used covariance function in GP modeling is the squared exponential 
covariance function with the set of hyperparameters~$\varphi=\{\lambda\in\R_+^*,\sigma_f\in\R_+\}$ 
\citep{rasmussen2006gaussian}:
\begin{align}
k_{\varphi}(\x,\x^\prime)=\sigma_f^2 \exp{\left(-\frac{\Vert \x- \x^\prime \Vert^2}{2\lambda^2} \right) }
\end{align}
The length-scale~$\lambda$ determines the 
number of expected upcrossing of the level zero in a unit interval by a 
zero-mean GP. The signal variance~$\sigma_f^2$ describes the average distance of 
the function~$f_{GP}(\x)$ away from its mean. With this kernel any realization of~$f_{GP}(\x)$ is a smooth function, which makes it a suitable candidate for modeling physical dynamics. \\
In this paper, we use GPs for multivariate regression. Since 
the output of a GP is one dimensional, a regression over~$n$ outputs
requires~$n$ GPs. Therefore, the vector valued function~$\bm 
m(\cdot)=[m_1(\cdot),\ldots,m_n(\cdot)]^\top$ describes the mean functions for 
each component of a vector-valued~${\bm f}_{GP}$. The covariance functions for each state are bundled 
in the function~$\bm 
k(\cdot,\cdot)=[k_{\varphi_1}(\cdot,\cdot),\ldots,k_{\varphi_n}(\cdot,\cdot)]^\top$ with
\begin{align}
f_{GP,i}(\x)\sim \GP(m_i(\x),k_{\varphi_i}(\x,\x^\prime))
\end{align}
for $i=1,\ldots,n$ and the corresponding set of hyperparameters~$\varphi_i$. The GP has to be 
provided with input/output pairs. For this purpose, we arrange the~$m$ training 
inputs~$\{\x_i\}_{i=1}^m$ and outputs~$\{\bm y_i\}_{i=1}^m$ pairs in an input
training matrix~$X=[\x_1,\x_2,\ldots,\x_m]$ and an output training matrix 
$Y=[\bm y_1,\bm y_2,\ldots,\bm y_m]^\top$. Therefore, the training data set for the 
$i$-th GP is described \mbox{by~$\mathcal D_i=\{X,Y_{:,i}\}$} where~$Y_{:,i}$ is the~$i$-th column of the matrix~$Y$. 
The hyperparameters~$\varphi_i$ are trained through likelihood optimization, 
thus by maximizing the probability of the seen data to occur given the current 
parameters and input values
\begin{align}
\varphi_i^* = \arg\max_{\varphi_i} \log P(Y_{:,i}|X,\varphi_i).
\end{align}
The predicted output~$\bm y^*\in\R^n$ for a test value~$ \x^*$ is a
Gaussian distributed variable.
With the assumption that the mean functions of the GPs are set to zero, a prediction of the~$i$-th component of~$\bm y^*\vert \x^*$ is obtained with
\begin{align}
y_i^* &\sim \mathcal{N} \left(\mean_i(\bm y^*), \var_i(\bm y^*)\right),\\
\label{for:meanvalue}
\mean_i(\bm y^*)&=\bm k_{\varphi_i}(\x^*,X)^\top (K_{\varphi_i}(X,X)+I_m {\sigma^2_{n}}_i)^{-1}Y_{:,i},\\
\var_i(\bm y^*)&=k_{\varphi_i}(\x^*,\x^*)-\bm k_{\varphi_i}(\x^*,X)^\top \notag\\
\label{for:varvalue}
& \phantom{{}=}(K_{\varphi_i}(X,X)+I_m {\sigma^2_{n}}_i)^{-1} \bm k_{\varphi_i}(\x^*,X),
\end{align}
where~$\mean_i(\cdot)$ is the mean and~$\var_i(\cdot)$ the variance of the 
random variable. The matrix $K_{\varphi_i}(X,X)$ denotes the concatenation of pairwise evaluation of all input data points and~$\bm k_{\varphi_i}(\x,X)$ the vector-valued extended covariance function with the set of hyperparameters~$\varphi_i$. 
The variable~${\sigma_{n}}_i^2\in\R_+$ is the variance of the input data 
for all~$i\in\{1,\ldots,m\}$. It increases the numerical stability of the 
matrix inversion. The~$n$ normally distributed components are combined in a 
multi-variable distribution
\begin{align}
\bm y^*&\sim \mathcal{N} \left( \bm \mean(\bm y^*), \var(\bm y^*)\right),\\
\bm \mean(\bm y^*)&=[\mean_1(\bm y^*),\ldots,\mean_n(\bm y^*)]^\top,\\
\Sigma(\bm y^*)&=\diag(\var_1(\bm y^*),\ldots,\var_n(\bm y^*)).
\end{align} 
\subsection{Dynamic Model of a Robot Manipulator}
The dynamics of an~$n$-link rigid manipulator can be written as
\begin{align}
\label{for:dyn_model_man}
H(\bm q)\ddot{\bm q}+C(\bm q,\dot{\bm q})\dot{\bm q}+\bm \g(\bm q)=\bm\tau,
\end{align}
where~$\bm q\in\R^n$ is the vector of continuous joint 
displacements and~$\bm\tau\in\R^n$ is the vector of applied joint torques. The 
matrix~$H(\bm q)\in\R^{n\times n}$ is the 
manipulator's inertia,~$C(\bm q,\dot{\bm q})\dot{\bm q}$ in~$\R^n$ is the vector 
of centripetal and Coriolis torques, and~$\g(\bm q)\in\R^n$ includes gravity
terms and other forces which act at the joints which have the following properties:
\begin{prope}[Structural properties]
\label{prop:structural}
\begin{itemize}
\item~$H(\bm q)$ is symmetric and positive definite.
\item~$\dot{H}(\bm q)-2C(\bm q,\dot{\bm q})\in\R^{n\times n}$ is a skew-symmetric matrix.
\end{itemize}
\end{prope}
Additionally, for robots with only rotational joints, the following properties 
hold~\citep{ghorbel1993positive}:
\begin{prope}[Boundedness and Linearity]
\label{prop:bounded}
\begin{itemize}
\item The inertia matrix is bounded and Lipschitz continuous, i.e.~$\Vert H(\bm q) \Vert<\infty$ and~$\Vert H(\q)-H(\q') \Vert\leq L \Vert \q-\q' \Vert$ with~$L>0$, for all~$\q,\q' \in \R^n$.
\item The matrix~$C(\bm q,\dot{\bm q})$ is bounded in~$\bm q$ and linear in~$\dot{\bm q}$, i.e.~$\Vert C(\bm q,\dot{\bm q}) \Vert\leq c_C\Vert \dot{\bm q} \Vert$ and~$C(\bm q,\dot{\bm q})\dot{\bm p}=C(\bm q,\dot{\bm p})\dot{\bm q}$ for all~$\bm q,\dot{\bm q},\dot{\bm p} \in\R^n$ and $c_C\in\R_+$.
\end{itemize}
\end{prope}
\subsection{Stochastic Stability Definitions}
Since we propose the use of GPs (a stochastic process), we quickly review the concept of stochastic stability and boundedness of the stochastic differential equation
\begin{align}
&d\x = \bm f(\x,t)dt + G(\x,t)d\bm{\omega}\label{eq:sde}\\
&\text{with } \bm f\colon\R^{n_1}\times\R_+\to\R^{n_1},\,G\colon\R^{n_1}\times\R_+\to\R^{n_1\times n_2},
\end{align}
where~$\bm{\omega}$ indicates the Brownian motion and $n_1,n_2\in\N$.
\begin{mydef}[Stochastic Sample Path Boundedness]
\label{def:SamplePathBoundedness}
Let there exist a ball~$B=\{\|\x\|\leq r| \x \in\R^{n_1},r>0\}$ and a time~$\tau_1$ 
denoting the first exit time from~$\R^{n_1}\backslash B$ for the solution~$\x(t)$ 
where~$\bm {x_0}\in\R^{n_1}\backslash B$.
The system \eqref{eq:sde} is \textit{stochastically sample path bounded}, if 
for each~$\varepsilon>0$ there exists a~$\delta>0$ such that 
\begin{align}
P\left(\sup_{0\leq t\leq \tau_1}\|\x(t)\| \leq \delta\right)>1-\varepsilon.
\end{align}
\end{mydef}
According to~\citet{gard1988introduction} this is shown as follows:
\begin{thm}
Let there exist a proper Lyapunov function~$V\in C^2$ for which the drift operator
\begin{align}
LV(\x) =\frac{\partial V}{\partial \x}\bm f
    +\frac{1}{2}\tr \left(G^\top \frac{\partial^2 V}{\partial \x \partial \x} G\right)\leq 0
\end{align}
holds for~$\x\in\R^{n_1}\backslash B$. Than the solution of~\eqref{eq:sde} 
is stochastic sample path bounded with~$B=\{\|\x\|\leq r| \x \in\R^{n_1},r>0\}$.
\end{thm}
A stronger stability criteria is given as follows:
\begin{mydef}[Stochastic Asymptotic Stability in the large]
\label{def:AsyStabilityLarge}
The system \eqref{eq:sde} is \textit{stochastically asymptotically
stable in the large} if it is stochastically stable in probability and for 
all~$\x_0\in\R^{n_1}$ holds
\begin{align}
P\left(\lim_{t\to\infty}\x(t) = \bm 0\right) = 1.
\end{align}
\end{mydef}
Stability in the large is shown as following \citep{mao2007stochastic}:
\begin{thm}
\label{theo:stochstability}
If there exists a positive-definite radially unbounded function~$V(\x)\in\C^{2}$
 such that the drift operator $LV$ is negative-definite, then the trivial solution of~\eqref{eq:sde} is 
stochastically asymptotically stable in the large.
\end{thm}
We will use these theorems to show that the tracking error is stochastically 
sample path bounded and for an infinite number of training points the tracking 
error is asymptotic stable in the large.  Since stability in the large 
considers all realization of the stochastic process, 
all conclusions also hold if the mean function of the control law is applied 
in a deterministic setting.
\section{Gaussian Processes for modified Computed-Torque}
\label{sec:ctrl_law}
The augmented PD control law from~\citet{siciliano2010robotics} requires 
knowledge 
about the model, i.e the matrices~$H,C$ and~$ \g$. To avoid this requirement, 
we use a modified version of the control law which learns the difference between 
the model, given by the estimates~$\hat H, \hat C$ and~$\hat{ \g}$ and the true 
robot dynamics. The estimates~$\hat H, \hat C,\hat{ \g}$ can be obtained from 
CAD data or from standard identification procedures.
\subsection{Conditions}
The goal is to design a control law~$\bm u_c$ which tracks the desired joint 
trajectory~$\bm{\q}_d,\bm{\dq}_d,\bm{\ddq}_d$ under the following conditions:
\begin{itemize}
\item[C1] The desired trajectory satisfies~$\Vert\qd\Vert<c_q$, 
           ~$\Vert\dqd\Vert<c_{\dot q}$, and~$\Vert\ddqd\Vert<c_{\ddot q}$
             with~$c_q,c_{\dot q},c_{\ddot q}\in\R_+$.
\item[C2] The feedback gain matrices~$K_d$ and~$K_p$ are positive definite
              and the smallest singular value of~$K_d$ is larger than~$\beta\in\R_+$, 
              thus~$\sigma_{min}(K_d)>\beta$.
\item[C3] The norm of the model error is affinely bounded by the norm of the 
            angular velocity, i.e.~$\Vert H(\q)\ddqd+C(\q,\dq)\dqd+ \g(\q)-\hat 
            H(\q)\ddqd-\hat C(\q,\dq)\dqd-\hat \g(\q)\Vert\leq \alpha+
            \beta\Vert \dq \Vert$ for all~$\q,\dq\in\R^n$ with~$\alpha,\beta\in\R_+$,
            and continuous regarding to~$\ddqd,\dqd, \dq, \q$.           
\end{itemize}
The conditions in C1, i.e. bounded reference motion trajectories, are a very 
natural assumption and do not pose any restriction in practice.
Also C2 does not restrict the applicability of our approach but must 
be kept in mind during the design of the controller. From a practical point of 
view, C3 states that the dynamics which are not modeled 
by~\eqref{for:dyn_model_man} can at most depend linearly on the joint 
velocity. If there is a known range of uncertainty only in the inertia 
parameter, the values $\alpha$ and $\beta$ can be computed using the approach 
of~\citet{takegaki1981new}. Since the payload is the major reason for the 
uncertainty, this approach is suitable for most application scenarios.
\subsection{Control Law}
\begin{figure}[b]
\begin{center}
\vspace{0.2cm}
	\begin{tikzpicture}[auto, node distance=3cm,>=latex']
\tikzstyle{block} = [draw, fill=white, rectangle, minimum height=2em, minimum width=1em]
\tikzstyle{sum} = [draw, fill=white, circle, node distance=1cm]
\tikzstyle{input} = [coordinate]
\tikzstyle{output} = [coordinate]
\tikzstyle{mid} = [coordinate]

    \node [input, name=input] {};
    
    \node [mid, right of=input,node distance=1cm] (mid_input) {};
    \node [block, right of=mid_input,yshift=-0.2cm,node distance=2cm] (controller) 
                                                    {$\scriptstyle \hat H(\q)\ddqd+\hat C(\dq,\q)\dqd+\hat G(\q)$};
    \node [block, above of=controller,node distance=1cm] (GP) {$\scriptstyle \bm f_{GP}(\ddqd,\dqd,\q)$};
    \node [sum, right of=controller,node distance=2.1cm] (sum_left) {};
    \node [sum, right of=sum_left,node distance=0.7cm] (sum_center) {};
    \node [block, above of=sum_center,node distance=1cm] (Kp) {$\scriptstyle -K_p$};
    \node [block, below of=sum_center,node distance=1cm] (Kd) {$\scriptstyle -K_d$};
    \node [sum, below of=Kd,node distance=0.8cm] (sum_Kd) {};
    \node [sum, above of=Kp,node distance=0.8cm] (sum_Kp) {};
    \node [block, right of=sum_center,node distance=1.3cm] (robot) {$\scriptstyle \text{manipulator}$};
    \node [output, above right=0.1cm and 0.8cm of robot.east] (q) {};
    \node [output, below right=0.1cm and 0.5cm of robot.east] (dq) {};
    \node [output, below right=0cm and 0.6cm of dq] (dqout) {};
    \node [mid, above of=sum_center,node distance=2.2cm] (mid_Kp) {};
    \node [mid, below of=sum_center,node distance=2.2cm] (mid_Kd) {};
    \node [input, left of=sum_Kp,xshift=-2.8cm] (input1) {};
    \node [mid, right of=input1,node distance=1.1cm] (mid_input1) {};
    \node [input, left of=controller,yshift=-0.2cm] (input2) {};
    \node [mid, right of=input2,node distance=0.8cm] (mid_input2) {};

	\draw [-] (input) -- node[above] {$ \scriptstyle \ddqd$} (mid_input);
	\draw [-] (input1) -- node[above] {$ \scriptstyle \qd$} (mid_input1);
	\draw [-] (input2) -- node[below] {$ \scriptstyle \dqd$} (mid_input2);
	
	\draw [->] (mid_input2) -- ([yshift=-0.2cm]controller.west);
	\filldraw (mid_input2) circle (2pt);
	\draw [->] (mid_input) -- ([yshift=0.2cm]controller.west);
	\draw [->] (mid_input2) |- ([yshift=0.1cm]GP.west);
	\draw [->] (controller) -- (sum_left);
	\draw [->] (sum_left) -- (sum_center); 
	\draw [->] (sum_center) -- node[] {$\scriptstyle \bm\tau$} (robot);
    \draw [->] (mid_input) |-  ([yshift=-0.1cm]GP.west);;
    \draw [->] (mid_input1) |- node[at end,xshift=-0.1cm] {-} (sum_Kp);
    \draw [->] (mid_input2) |- node[at end,xshift=-0.1cm] {-} (sum_Kd);
    \draw [->] (GP) -| node {} (sum_left);
    \draw [-] ([yshift=0.1cm]robot.east) -| node[above,xshift=-0.6cm] {$\scriptstyle \q$} (q);
    \draw [-] ([yshift=-0.1cm]robot.east) -| node[below,xshift=-0.3cm] {$\scriptstyle \dq$} (dq);
    
    \draw [->] (q) |- (mid_Kp) -| ([xshift=0.3cm]GP.north);
    \draw [->] (dq) |- (mid_Kd) -| ([xshift=0.3cm]controller.south);

    \draw [->] (q) |- ([yshift=-0.3cm]mid_Kd.center) -| ([xshift=-0.3cm]controller.south);
   
    \draw [->] (mid_Kp) -- (sum_Kp);
    \draw [->] (sum_Kp) -- (Kp);
    \draw [->] (Kp) -- (sum_center);
	
    \draw [->] (mid_Kd) -- (sum_Kd);
    \draw [->] (sum_Kd) -- (Kd);
    \draw [->] (Kd) -- (sum_center);

\end{tikzpicture}
	\vspace{-0.5cm}\caption{Structure of the proposed closed loop 
	controller~\eqref{for:control_law}.}
	\label{fig:CL}
\end{center}
\end{figure}
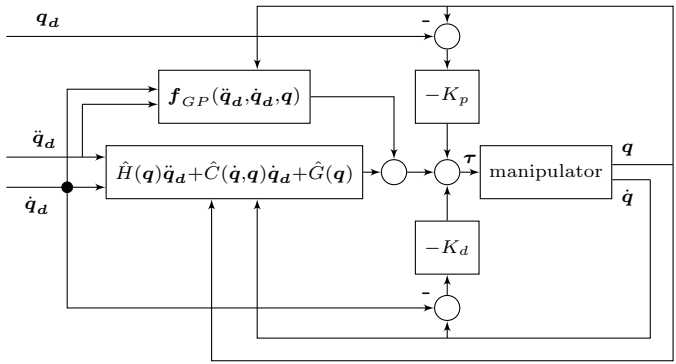

The following theorem proposes a control law to ensure a bounded tracking 
error under the proposed conditions.
\begin{thm}
\label{theo:ctrl_law}
Assume an~$n$-link rigid manipulator with only rotational joints 
\begin{align}
\label{eq:DynManipulator}
 \bm\tau=H(\q)\ddq+C(\q,\dq)\dq+\g(\q),
\end{align} 
for which Properties~\ref{prop:structural} and \ref{prop:bounded} and C1-C3 hold. 
Given an estimated model of the manipulator
\begin{align}
\hat{\bm\tau}=\hat H(\q)\ddq+\hat C(\q,\dq)\dq+\hat \g(\q)
\end{align}
with control law
\begin{align}
\bm{u_c}&=\hat H(\q)\ddqd+\hat C(\q,\dq)\dqd+\hat \g(\q)\notag\\
&+\bm f_{GP}(\qc)-K_d \dqe-K_p \qe\label{for:control_law},
\end{align}
shown in Fig.~\ref{fig:CL}, where~$\qc=[\ddqd^\top, \dqd^\top, \q^\top]^\top$ and~$\bm f_{GP}$ contains~$n$ 
posteriors of GPs with squared exponential kernel
\begin{align}
\bm f_{GP}(\qc)=\begin{bmatrix}
f_{GP,1}\sim \GN(\mu_1(\qc),\var_1(\qc))\\
\vdots\\
f_{GP,n}\sim \GN(\mu_n(\qc),\var_n(\qc))\\
\end{bmatrix}.
\end{align} 
Then, the tracking error~$\qe=\q-\qd$ is stochastically sample path bounded.
\end{thm}
The stability proof of the control law proposed in Theorem~\ref{theo:ctrl_law} 
bases on the work in~\citet{whitcomb1993comparative}.
\begin{pf}
With~$\bm\tau=\bm{u_c}$ the closed loop system is given by
\begin{align}
\ddqe&=H(\q)^{-1}(\hat H(\q)\ddqd+\hat C(\q,\dq)\dqd-C(\q,\dq)\dq+\hat \g(\q)\notag\\
&-\g(\q)+\bm f_{GP}(\qc)-K_d \dqe-K_p \qe)-\ddqd,
\end{align}
since the matrix~$H(\q)$ is always non-singular. The posterior of the
 GP~$\bm f_{GP}(\qc)$ can be split in a drift~$\bm \mu(\qc)$ and a 
 diffusion term~$\Sigma(\qc)\bm w$
\begin{align}
\bm f_{GP}(\qc)&=\bm \mu(\qc)+\Sigma(\qc)\bm w, \label{for:drift_diff}\\
\bm \mu(\qc)&=\begin{bmatrix}
\bm k_{\varphi_1}^\top (K_{\varphi_1}+I_m {\sigma^2_{n}}_1)^{-1} Y_{:,1}\\
\vdots\\
\bm k_{\varphi_n}^\top (K_{\varphi_n}+I_m {\sigma^2_{n}}_n)^{-1} Y_{:,n}\\
\end{bmatrix}, \nonumber \\
\Sigma(\qc)&=\diag\begin{bmatrix}
k_{\varphi_1}-\bm k_{\varphi_1}^\top (K_{\varphi_1}+I_m {\sigma^2_{n}}_1)^{-1} \bm k_{\varphi_1}\\
\vdots\\
k_{\varphi_n}-\bm k_{\varphi_n}^\top (K_{\varphi_n}+I_m {\sigma^2_{n}}_n)^{-1} \bm k_{\varphi_n}
\end{bmatrix}^{\frac{1}{2}}, \nonumber
\end{align}
with an~$n$-dimensional standard Brownian noise vector~\mbox{$\bm w=[w_1 \dots w_n]^\top$}. We reformulate the closed loop system with a drift and a diffusion term 
\begin{align}
\frac{d}{dt} \begin{bmatrix} \dqe\\ \vphantom{e} \\ \qe \end{bmatrix}&=\begin{bmatrix}
 H^{-1}(\hat H\ddqd+\hat C\dqd-C\dq+\hat \g-\g\\
+\bm \mu(\qc)-K_d \dqe-K_p \qe)-\ddqd\\
\dqe
\end{bmatrix}\notag\\
&+\begin{bmatrix} H^{-1}\Sigma(\qc) \\ 0  \end{bmatrix}\bm w:=\bm f(\bm\xi,t) + G(\bm\xi) \bm w, \label{for:sde}
\end{align}
where the vector~$\bm\xi=[\q^\top,\dq^\top]^\top \in \X \subset \R^{2n}$.
For the stability analysis of this stochastic differential equation we use the differential generator~$L$ which maps~$C^2$ functions~$V \colon \X\to \R$ to~$C^0$ functions~$LV \colon \X\to \R$ given by Theorem~\ref{theo:stochstability}.
Assume the following Lyapunov function
\begin{align}
V(\bm\xi)=\underbrace{\frac{1}{2}\dqe^\top H \dqe+\frac{1}{2}\qe^\top K_p 
\qe}_{V_0(\bm\xi)}+\underbrace{\vphantom{\frac{1}{2}}\varepsilon \qe^\top H 
\dqe}_{V_{\text{mix}}(\bm\xi)},\label{for:lyap}
\end{align}
which contains the two terms~$V_0$ and~$V_{\text{mix}}$. We start with the 
computation of~$LV_0$ using the property that the matrix~$H^{-1}$ is symmetric
\begin{align}
LV_0&=\dqe^\top H \ddqe+\frac{1}{2}\dqe^\top \dot H \dqe+\dqe^\top K_p \qe\notag\\
&+\frac{1}{2}\tr \Sigma^\top(\qc) H^{-1} \Sigma(\qc).
\end{align}
Now, the acceleration error~$\ddqe$ of~\eqref{for:sde} is substituted.
\begin{align}
LV_0&=\dqe^\top(\hat H\ddqd+\hat C\dqd-C\dq+\hat \g-\g-K_d \dqe\notag\\
&-K_p \qe-H\ddqd+\bm \mu(\qc))+\dqe^\top\left(\frac{1}{2}( \dot H-2C)+C\right) \dqe \notag\\
&+\dqe^\top K_p \qe+\frac{1}{2}\tr \Sigma^\top(\qc) H^{-1} \Sigma(\qc).
\end{align}
Since Property~\ref{prop:structural} implies the skew-symmetry of~$\dot H-2C$, this term can be canceled out and~$LV_0$ is given by
\begin{align}
LV_0&=-\dqe^\top K_d \dqe + \dqe^\top(\tilde H\ddqd+\tilde C\dqd+\tilde \g+\bm \mu(\qc))\notag\\
&+\frac{1}{2}\tr \Sigma^\top(\qc) H^{-1} \Sigma(\qc),\label{for:V0}
\end{align}
with the difference between the manipulator model and the estimated model 
matrices defined by 
\begin{align}
\tilde H(\q)&=\hat H(\q)-H(\q),\nonumber \\
\tilde C(\q,\dq)&=\hat C(\q,\dq)-C(\q,\dq),\label{eq:modeldev}\\
\tilde \g(\q)&=\hat \g(\q)-\g(\q). \nonumber
\end{align}
The first summand of $LV_0$ is negative definite but the definiteness of the second part 
is indeterminate. However, condition C3 guarantees that~$\Vert\tilde 
H\ddqd+\tilde C\dqd+\tilde \g\Vert$ is bounded by an affine function 
$\alpha+\beta\Vert\dq\Vert$.  The mean prediction~$\Vert \bm \mu(\qc)\Vert$ and the 
corresponding variance~$\Vert \Sigma(\qc)\Vert$ is also 
bounded~\citep{beckers:cdc2016}. Therefore, it is possible to find an upper 
bound for the 
second and third part of~\eqref{for:V0}
\begin{align}
&\Vert \dqe^\top(\tilde H \ddqd+\tilde C\dqd+\tilde \g+\bm \mu(\qc))\Vert
\hspace{-0.1cm}\leq\hspace{-0.1cm} \Vert\dqe\Vert (\alpha+c_{\mu}+\beta\Vert\dq\Vert).\hspace{0.3cm}\label{for:LV02}
\end{align}
Since~$\Vert H(\dq)\Vert$ is bounded and the matrix is always non-singular, the inverse~$\Vert H^{-1}(\dq)\Vert$ is also bounded
\begin{align}
\frac{1}{2}\tr \Sigma^\top(\qc) H^{-1} \Sigma(\qc) \leq c_g\in\R_+.
\end{align}
These results are used for the estimation of an upper bound for~$LV_0$
\begin{align}
LV_0\leq-\dqe^\top K_d \dqe +\Vert\dqe\Vert\beta\Vert\dqe\Vert
+\Vert\dqe\Vert (\alpha+\beta c_{\dot q}+c_{\mu})+c_g.\hspace{0.5cm}\label{for:LV0}
\end{align}
If~$\sigma_{min}(K_d)>\beta$, the first part dominates the second part of the equation and the sum is negative definite.
Since it is negative definite, the quadratic part of~$LV_0$, i.e.~$-\dqe^\top K_d \dqe+\Vert\dqe\Vert\beta\Vert\dqe\Vert$, dominates the linear and the constant part for~$\dqe\to\infty$. Thus the following holds
\begin{align}
\lim_{\Vert \dqe \Vert\to\infty} L V_0(\bm\xi,t)=-\infty,
\end{align}
for all~$\qe\in\R^n$. With the continuity of~$V_0$, there exist a 
ball~$B_{\dot{\bm e}}=\{\Vert \dqe \Vert \leq \delta_{\dot e}\}$ with the property that~$ 
LV_0(\bm\xi,t)<0$ if~$\dqe\in\R^n\setminus B_{\dot{\bm e}}$. In other words,~$LV_0$ 
is negative outside~$B_{\dot{\bm e}}$ and therefore $\dqe$ is stochastically sample path bounded. Boundedness of the tracking error $\qe$ can not be guaranteed 
so far because~$LV_0$ does not depend on~$\qe$. Therefore, the second part 
$V_{\text{mix}}$ of Lyapunov function~\eqref{for:lyap} is included
\begin{align}
LV_{\text{mix}}&=\varepsilon(\dqe^\top H \dqe+\qe^\top \dot H \dqe+\qe^\top H 
\ddqe)\\
&=\varepsilon \dqe^\top H \dqe+\varepsilon \qe^\top \dot H \dqe+\varepsilon \qe^\top( -K_p \qe - K_d\dqe\notag\\
&+\tilde H\ddqd+\tilde C\dqd-C\dqe+\tilde \g+\bm \mu(\qc)).
\end{align}
After rewriting the equation
\begin{align}
LV_{\text{mix}}&=\varepsilon \Big(\underbrace{-\qe^\top K_p \qe}_{LV_1}+
\underbrace{\vphantom{K_p}\dqe^\top H \dqe}_{LV_2}+
\underbrace{\qe^\top (\dot H-C-K_d) \dqe}_{LV_3}\notag\\
&+\underbrace{\qe^\top\left(\tilde H\ddqd+\tilde C\dqd+\tilde \g+\bm \mu(\qc)\right)}_{LV_4}\Big),
\end{align}
the term is analyzed: The first summand~$LV_1$ is negative 
definite from C2. The second part~$LV_2$ sums up with~$-\dqe^\top K_d\dqe$ of~$LV_0$, 
see~\eqref{for:LV0}, but since~$\varepsilon$ can be arbitrary small the sum is 
still negative definite. For the cross term~$LV_3$ it is sufficient to show the 
boundedness of the operator norm of~$(\dot H - C-K_d)$. The upper bound
\begin{align}
LV_3\leq \left(\Vert K_d \Vert +\frac{5}{2} \left\Vert \frac{\partial H}{\partial \q}\right\Vert \Vert \dq \Vert \right) \Vert \qe \Vert \Vert \dqe \Vert\leq c_{V3} \Vert \qe \Vert \Vert \dqe \Vert ,
\end{align}
with~$c_{V3}\in\R_+$ can be shown by using the chain rule and the dependency between $\dot H$ and $C$. The partial 
derivation~$\left\Vert \frac{\delta H}{\delta \q}\right\Vert$ is a bounded 
operator since~$H(\q)$ is Lipschitz continuous. So~$LV_3$ is 
a~$\varepsilon$-size bounded operator on~$\qe$ and~$\dqe$ which preserve the 
negative definiteness. The last part~$LV_4$ is bounded by
\begin{align}
LV_4&\leq \Vert\qe\Vert\beta\Vert\dqe\Vert+\Vert\qe\Vert \underbrace{(\alpha+\beta c_{\dot q+}c_{\mu})}_{c_{V4}}
\end{align}
with~$c_{V4}\in\R_+$ which is analog to~\eqref{for:LV02}. After combining the parts, the upper bound for the diffusion operator~$L$ of the overall Lyapunov function~$V$ is given by:
\begin{align}
LV&=-\dqe^\top (K_d-\varepsilon H) \dqe-\varepsilon\qe^\top K_p \qe +\varepsilon\qe^\top (\dot H- C-K_d) \dqe\notag\\
&+(\dqe^\top+\varepsilon\qe^\top)(\tilde H\ddqd+\tilde C\dqd+\tilde \g+\bm \mu(\qc))\notag\\
&+\frac{1}{2}\tr \Sigma^\top(\qc) H^{-1} \Sigma(\qc)\label{for:LV}\\
&\leq -\dqe^\top (K_d-\varepsilon H)\dqe +\Vert\dqe\Vert\beta\Vert\dqe\Vert+(c_{V4}+c_g)\Vert\dqe\Vert\notag\\
&\phantom{\leq}-\varepsilon\qe^\top K_p \qe+\varepsilon (c_{V3}+\beta) \Vert \qe \Vert \Vert \dqe \Vert+\varepsilon c_{V4} \Vert\qe\Vert. \label{for:LVUP}
\end{align}
In comparison to~\eqref{for:LV0}, this term also includes a dominant quadratic part which depends on the error~$\qe$. Therefore, it is possible to find an~$\varepsilon>0$ which creates the ball 
\begin{align}
B_{\bm \xi}&=\{\Vert \bm\xi\Vert \leq \delta_{\dot e e}\}\\
\text{with } LV&<0 \text{ for } \bm\xi\in\R^{2n}\setminus B_{\bm \xi}.\notag
\end{align}
Consequently, the tracking error is stochastically sample path bounded and 
enters in a finite time the set~$B_{\bm \xi}$. 
\hfill \rule{1.5ex}{1.5ex}	
\end{pf}
\begin{myremark}
The proposed control law uses Gaussian Processes with zero mean functions which is reasonable if no prior knowledge about the model error is given. However, if a priori knowledge is available, the Gaussian Process regression can be supported by a nonzero mean function. A bounded tracking error is preserved if (\ref{for:LV02}) remains bounded which is fulfilled as long as the mean function is bounded~\citep{beckers:cdc2016}.
\end{myremark}
The proof applies Theorem~\ref{theo:stochstability}
to show that the tracking error is stochastically sample path bounded with the ball~$B_{\bm \xi}$. 
For a radially unbounded, positive-definite Lyapunov function, it shows that the 
drift operator is negative outside of this ball and therefore the tracking error enters the ball in a finite time.\\
Equation \eqref{for:V0} shows the need of C3 to ensures the global negative definiteness of $LV_0$ outside the ball. However, with less restrictions on the model error, it is possible to find local areas of boundedness which requires appropriate initial states.
Important to note here is the stochastic nature of the control low~$\bm{u_c}$ as it
uses with~$\bm f_{GP}$ a stochastic process. Nevertheless, its deterministic 
counterpart (the mean function) also results in the desired property:
\begin{cor}
\label{Cor:meanstable}
The rigid manipulator described by~\eqref{eq:DynManipulator} with control law \eqref{for:control_law} results in a bounded tracking error if the stochastic process~$\bm f_{GP}(\qc)$ is replaced by its deterministic posterior mean function $\bm \mean(\qc)$ as defined in~\eqref{for:meanvalue}.
\end{cor}
This corollary directly follows from Theorem~\ref{theo:ctrl_law} since with  
stochastic sample path boundedness it must hold for all realizations of the 
stochastic control law. The mean function is simply one of these realizations.

\subsection{Training}
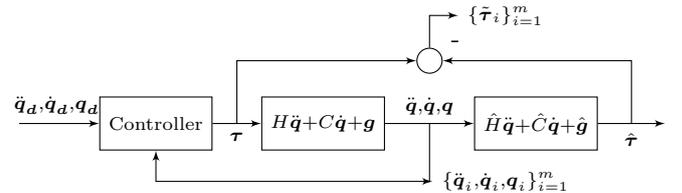
\begin{figure}[b]
\begin{center}
 \begin{tikzpicture}[auto, node distance=3cm,>=latex']
\tikzstyle{block} = [draw, fill=white, rectangle, minimum height=2em, minimum width=1em]
\tikzstyle{sum} = [draw, fill=white, circle, node distance=1cm]
\tikzstyle{input} = [coordinate]
\tikzstyle{output} = [coordinate]
\tikzstyle{t_output} = []
\tikzstyle{pinstyle} = [pin edge={to-,thin,black}]

    \node [input, name=input] {};
    \node [block, right of=input,node distance=1.8cm] (controller) {$\scriptstyle \text{Controller}$};
    \node [block, right of=controller,node distance=2.2cm] (robot) {$\scriptstyle H\ddq+C\dq+\bm\g$};
    \node [block, right of=robot,node distance=2.8cm] (hatrobot) {$\scriptstyle \hat H\ddq+\hat C\dq+\hat{\bm g}$};
    
    \node [output, right of=hatrobot,node distance=1.7cm] (output) {};

	\draw [->] (input) -- node[name=t1] {$\scriptstyle \ddqd,\dqd,\qd\:$} (controller);
	\draw [->] (controller) -- node[name=t,below] {$\scriptstyle \bm\tau$} (robot);
    \draw [->] (robot) -- node[name=q] {$\scriptstyle \ddq,\dq,\q$} (hatrobot);
    \draw [->] (hatrobot) -- node[name=o, below] {$\scriptstyle \hat{\bm\tau}$} (output);
    
    \coordinate [below of=q,node distance=1cm] (dp) {};
    \node [t_output, right of=dp,node distance=1cm] (x_output) {$\scriptstyle\{\ddq_i,\dq_i,\q_i\}_{i=1}^m$};
    \node [t_output, above of=x_output,node distance=2.2cm] (y_output) {$\scriptstyle\{\tilde{\bm \tau}_i\}_{i=1}^m$};
    \draw [->] (q) |- (dp) -| (controller);

    \node [sum, above of=q,node distance=0.6cm,label={[label distance=0cm]20:-}] (sum) {};
    
    \draw [->] (dp) -- (x_output);
    \draw [->] (sum) |- (y_output);
    
    \draw [->] (t) |- (sum);
    \draw [->] (o) |- (sum);

\end{tikzpicture}
	\vspace{-0.5cm}\caption{The structure for generating the training data 
	set~$\mathcal D=\{[\ddq_i,\dq_i,\q_i],\tilde{\bm\tau_i}\}_{i=1}^{m}$ for the GPR.\label{fig:TD}}
\end{center}
\end{figure}

While the previous section proposed the control law which guarantees 
stochastic sample path boundedness around the origin, this section shows how 
this can tend to asymptotic stability through training of the GP. 
So practically speaking, training aims to shrink the radius of the ball~$B_{\bm 
\xi}$ as the number of training points increases. For this purpose, the 
Gaussian Process learns the difference between the real and the estimated 
dynamics of the manipulator
\begin{align}
\label{for:learn}
\tilde{\bm \tau} =\bm \tau-\hat{\bm \tau} = - \tilde H(\q)\ddq-\tilde C(\q,\dq)\dq-\tilde \g(\q).
\end{align} 
Figure~\ref{fig:TD} shows how to generate training pairs of various joint states~$\{\ddq_i,\dq_i,\q_i\}_{i=1}^m$ 
and differences between the applied torque and the estimated 
torque of the model~$\{\tilde{\bm \tau}_i\}_{i=1}^m$. The manipulator can be excited directly 
with a torque or with a stabilizing controller to drive the manipulator in the desired states. 
It is advisable that the area of training points is similar to the desired operation area of the manipulator. An appropriate choice of training points inside this area can be done, for example, with the Bayesian optimization method where the next training point is set to the position of maximum variance, as proposed in \citet{sui2015safe}. After recording the training pairs, 
the Gaussian Process can be trained using likelihood optimization.\\
To achieve a asymptotically stable tracking error, the mean function~$\bm \mu(\qc)$ of 
the GP must cancel out the model error~$ \tilde{\bm \tau}$ 
given in~\eqref{for:learn}. We introduce the following lemma for the analysis of the convergence.
\begin{mylemma}[Consistency]
Let~$\bm f_0:\R^n\to\R^n$ a continuous function. A set of~$m$ training points 
is distributed uniformly in a bounded space or in a stochastic way with a nonzero density function.\\
A Gaussian
Process~$\bm f_{GP}$ with squared exponential kernel is consistent, i.e.
\begin{align}
\label{eq:consistency}
\expval_{f_0} \left\| \expval\left( \bm f_{GP}\right)-\bm f_0\right\|^2
\stackrel{m\to\infty}{\longrightarrow} 0,
\end{align}
on the compact domain \citep{vaart2011information}.
\end{mylemma}
As the continuity condition is fulfilled through assumption C3, the difference
between the model error and the GP posterior tends 
to zero as the number of training points approaches infinity 
which incorporates that the variance is also zero.
This allows the following conclusion:
\begin{cor}
\label{cor:asymptotic}
The tracking error for the rigid manipulator~\eqref{eq:DynManipulator} 
with control law~\eqref{for:control_law} is asymptotically stable in the large 
as the number of training points for the GP approaches infinity.
\end{cor}
\begin{pf}
\label{proof:asymptotic}
If the number of training points tends to infinity, lemma~\ref{eq:consistency} shows that 
\begin{align}
\label{for:learn1}
\bm \mean(\qc)=-\tilde H(\q)\ddqd-\tilde C(\q,\dqd)\dqd-\tilde \g(\q)
\end{align}
holds. The upper bound for the drift operator of the Lyapunov function~\eqref{for:LV} can now be rewritten as
\begin{align}
LV\leq&-\dqe^\top (K_d-\varepsilon H) \dqe-\varepsilon\qe^\top K_p \qe +\varepsilon\qe^\top (\dot H-C- K_d) \dqe\notag\\
&+\Vert\dqe\Vert\beta\Vert\dqe\Vert+\varepsilon\qe^\top\tilde C(\q,\dqd)\dqe.
\end{align}
With C2, the tracking 
error is asymptotically stable in the large. Thus for the limit value consideration of infinite many training points, the tracking error will 
converge to zero for any~$[\dq_0^\top,\q_0^\top]^\top\in\X$. 
\hfill \rule{1.5ex}{1.5ex}	
\end{pf}
\begin{myremark}
The GP is trained over $\ddq,\dq,\q$ but receives $\ddqd, \dqd, \q$ as inputs in the control law. This 
is beneficial for practical implementation as then no feedback of the manipulator's acceleration and the velocity is required.
Additionally, the dependency of~$\tilde C$ on~$\dq$ is problematic because it cannot be 
identified isolated from the the angular velocity~$\dq$ which is multiplied with~$\tilde C$. Therefore,
we use the desired velocity for the input of the Gaussian Process.
\end{myremark}


\section{Simulation and Experiment}
\label{sec:SimandExp}
\subsection{Simulation}
As unmanned aerial vehicles (UAVs) increasingly gain importance in automation 
and robotics, for simulation, we consider a model of a NACA-0015 airfoil moved 
through the air as illustrated in Fig.~\ref{fig:wing}. The inertia~$J_a$ of the 
wing is assumed as~$\SI{1}{\kilogram\meter^2}$, the mass~$m=\SI{1}{\kilogram}$ 
and the distance between the joint and the center of mass~$l=\SI{1}{\meter}$ . 
The goal is to control the angle~$q$ of the wing with an input torque~$\tau$. 
The wing is affected by an aerodynamic force which can be decomposed in lift and 
drag. \\
These forces depend on the angle of attack which is the angle between the 
direction of the air flow and the reference line of the wing. For a large angle 
of attack the lift and drag force are highly nonlinear and difficult to model 
mathematically since air flow becomes turbulent.
Our simulations are based on the measurements of wing taken in a wind 
tunnel~\citep{sheldahl1981aerodynamic}. For the model shown in 
Fig.~\ref{fig:wing}, the lift and drag forces are converted in the resulting 
torque and gravity is added.

Assume a damping free pendulum for the estimated dynamics
\begin{align}
\hat{J}_a\ddot{q}+\hat{m}g\hat{l}\sin(q)=\hat \tau,
\end{align}
with the estimated parameters~$\hat{J}_a=0.9J_a$, \mbox{$\hat{l}\hat{m}=0.9ml$.} 
Figure~\ref{fig:control_mod} shows the simulation results for the classical augmented PD 
control law using the estimated model. The feedback terms  are set to 
$K_p=K_d=5$ and the desired trajectory (dashed) is sinusoidal.
\begin{figure}[tb]
\centering
\newcommand{\nvar}[2]{%
    \newlength{#1}
    \setlength{#1}{#2}
}

\nvar{\dg}{0.3cm}
\def\dw{1}\def\dh{3}
\nvar{\ddx}{1.5cm}

\def\joint{%
    \filldraw [fill=white] (0,0) circle (5pt);
    \fill[black] circle (2pt);
}
\def\grip{%
    \draw[ultra thick](0cm,\dg)--(0cm,-\dg);
    \fill (0cm, 0.5\dg)+(0cm,1.5pt) -- +(0.6\dg,0cm) -- +(0pt,-1.5pt);
    \fill (0cm, -0.5\dg)+(0cm,1.5pt) -- +(0.6\dg,0cm) -- +(0pt,-1.5pt);
}
\def\robotbase{%
	\begin{scope}[shift=(0:0), rotate=90]
    \draw[rounded corners=8pt] (-\dw,-\dh)-- (-\dw, 0) --
        (0,\dh)--(\dw,0)--(\dw,-\dh);
    \end{scope}
}

\newcommand{\angann}[2]{%
    \begin{scope}[red]
    \draw [dashed, red] (0,0) -- (1.2\ddx,0pt);
    \draw [->] (\ddx,0pt) arc (0:#1:\ddx);
    \draw [->,black] (#1:\ddx) -- node [left] (a) {lift} +(0,\ddx/2);
    \draw [->,black] (#1:\ddx) -- +(\ddx/2,0) node [right] (a) {drag};
    \node at (#1/2-2:\ddx+8pt) {#2};
    \end{scope}
}

\newcommand{\torqueann}[4]{%
    \begin{scope}
    \draw [<-]([shift={(0,0)}]#1:10pt) arc[radius=10pt, start angle=#1, end angle=#2];
    \node at (#1/2+#2/2:15pt) {#3};
    \end{scope}
}

\newcommand{\lineann}[4][0.5]{%
    \begin{scope}[rotate=#2, blue,inner sep=2pt]
        \draw[dashed, blue!40] (0,0) -- +(0,#1)
            node [coordinate, near end] (a) {};
        \draw[dashed, blue!40] (#3,0) -- +(0,#1)
            node [coordinate, near end] (b) {};
        \draw[|<->|] (a) -- node[fill=white] {#4} (b);
    \end{scope}
}

\newcommand{\link}[1]{%
					\draw[scale=3,rotate=#1] node[left=0cm] {} plot file{data/naca0015.dat} -- cycle;
					 \draw[scale=3,rotate=#1-90,dashed] (0,0) -- (0,1)

}

\def\thetaone{30}

\begin{tikzpicture}
    \draw[->, >=latex, draw=gray, line width=3pt](-2,1.3)-- (-1,1.3);
    \draw[->, >=latex, draw=gray, line width=3pt](-2,0.5)-- (-1,0.5);
    \draw[->, >=latex, draw=gray, line width=3pt](-2,-0.3)-- (-1,-0.3);
    \draw [->, draw=gray, line width=2pt]([shift={(3,1.2)}]90:0.3cm) arc[radius=0.3cm, start angle=90, end angle=-180];
    \begin{scope}[shift=(0:0), rotate=0]
    \angann{\thetaone}{$q$};
    \torqueann{270}{160}{$\tau$};
    \link{\thetaone};
    \joint
    \end{scope}
    \draw[->] (4cm,1cm) -- node[right] {$g$} (4cm,0cm);
\end{tikzpicture}
    \caption{Model of torque controlled wing. Lift / drag forces 
                are highly nonlinear functions of the angle of attack~$q$.}
     \label{fig:wing}
\end{figure}
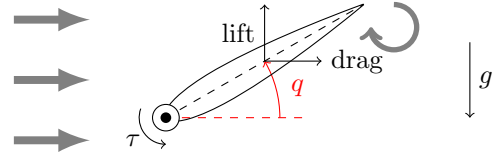
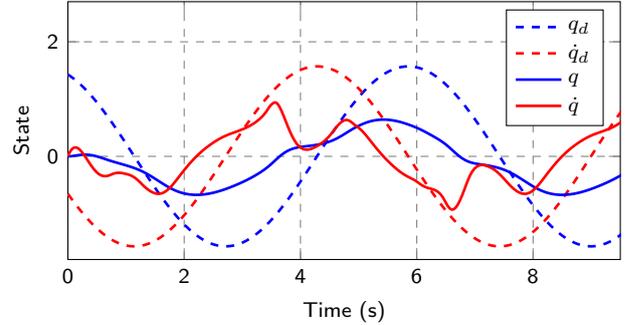
\begin{figure}[tb]
\begin{tikzpicture}
\begin{axis}[
  xlabel={Time (s)},
  ylabel={State},
  legend pos=north east,
  grid style={dashed,gray},
  grid = both,
  width=\columnwidth,
  height=5cm,
  ymin=-1.8,
  ymax=2.7,
  xmin=0,
  xmax=9.5]
\addplot[color=blue, dashed, line width=1pt] table [x index=0,y index=1]{data/figure2_model.dat};
\addplot[color=red, dashed, line width=1pt] table [x index=0,y index=2]{data/figure2_model.dat};
\addplot[color=blue,line width=1pt] table [x index=0,y index=3]{data/figure2_model.dat};
\addplot[color=red,line width=1pt] table [x index=0,y index=4]{data/figure2_model.dat};
\legend{$q_d$,$\dot{q}_d$,$q$,$\dot{q}$};
\end{axis}
\end{tikzpicture} 
      \caption{Classical augmented PD control with an estimated model 
            does not generate satisfactory results in comparison to the proposed control law which is shown in Fig.~\ref{fig:control_gp}. The dashed lines 
            are the desired joint position and velocity and the solid lines 
            show the true values. }
\label{fig:control_mod}
\end{figure}
Since the model contains parameter imprecision and influence of the airflow is 
not covered, the joint angle~$q$ differs from the desired~$q_d$.

In the next step, the proposed control law~\eqref{for:control_law} is
used. First, the GP learns the difference between the 
estimated model and the real wing. For this purpose, we generate~$990$ 
homogeneous 
distributed pairs of torques~$\tau$ and initial positions~$q_0$ on the set 
$[-8,8]\times[-\pi,\pi]$ to generate training points as shown in 
Fig.~\ref{fig:TD}. The initial joint velocity and acceleration is set to zero. 
In this example, we do not use an extra controller but apply the torque directly 
for a short time interval to the manipulator. After~$\SI{0.5}{\second}$ the 
joint position, velocity and acceleration~$\{\ddot{q},\dot{q},q\}$ are recorded. 
These values are inserted into the model to compute the estimated torque~$\hat 
{\tau}$. The difference between the applied torque and the estimated torque 
$\tilde {\tau}=\tau-\hat {\tau}$ is saved. The values~$\{\ddot{q},\dot{q},q\}$ 
and~$\{\tilde {\tau}\}$ build up a training pair.

\begin{figure}[tb]
\centering
\begin{tikzpicture}
\begin{axis}[
  xlabel={Time (s)},
  ylabel={State},
  legend pos=north east,
  grid style={dashed,gray},
  grid = both,
       width=\columnwidth,
  height=5cm,
  ymin=-1.9,
  ymax=2.6,
  xmin=0,
  xmax=9.3]
\addplot+[name path=varp1, color=gray,opacity=0.3, no marks] table [x index=0,y expr=\thisrowno{1}+\thisrowno{3}]{data/figure2_model_GP.dat};
\addplot+[name path=varm1, color=gray,opacity=0.3, no marks] table [x index=0,y expr=\thisrowno{1}-\thisrowno{3}]{data/figure2_model_GP.dat};
\addplot[gray,opacity=0.5] fill between[ of = varm1 and varp1]; 
\addplot+[name path=varp3, color=gray, no marks] table [x index=0,y expr=\thisrowno{2}+\thisrowno{4}]{data/figure2_model_GP.dat};
\addplot+[name path=varm3, color=gray, no marks] table [x index=0,y expr=\thisrowno{2}-\thisrowno{4}]{data/figure2_model_GP.dat};
\addplot[gray,opacity=0.5] fill between[ of = varm3 and varp3];
\addplot[color=blue,dashed, line width=1pt] table [x index=0,y index=1]{data/figure2_model.dat};
\addplot[color=red,dashed, line width=1pt] table [x index=0,y index=2]{data/figure2_model.dat};
\addplot[color=blue,line width=1pt] table [x index=0,y index=1]{data/figure2_model_GP.dat};
\addplot[,color=red,line width=1pt] table [x index=0,y index=2]{data/figure2_model_GP.dat};
\legend{,,,,,,$q_d$,$\dot{q}_d$,$q$,$\dot{q}$};
\end{axis}
\end{tikzpicture} 
      \vspace{-0.2cm}\caption{The proposed GP-based control law strongly reduces the tracking error in comparison to the classical augmented PD control. The mean (solid line) of the joint angle/velocity converges to a 
                  tight bound around the desired trajectory (dashed line). The 
                  shaded area marks the~$2\sigma$ interval of the $1000$ 
                  simulations.}
      \label{fig:control_gp}
\end{figure}
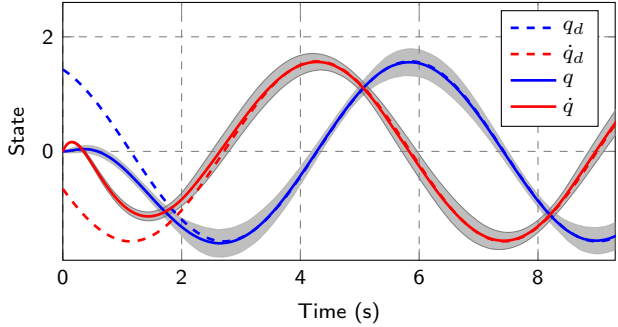

The GP is trained by this collection of training pairs and the 
hyperparameters of the squared exponential covariance function are optimized with 
a gradient method. Afterwards, the proposed control law~\eqref{for:control_law}
with the same desired trajectory and feedback gains is used. 
To show the effect of the stochastic control law, we simulate~1000 realizations
of the stochastic differential equation with a sample time of~$\SI{1}{\milli\second}$.
Figure~\ref{fig:control_gp} shows the mean (solid) and standard deviation 
(gray area) of the joint angle/velocity and the desired angles/velocity (dashed). The stochastic behavior is based on the stochastic prediction of the GP since the finite number of training data generates only an uncertain model.
Since the GP cancels the 
uncertainties of the model, the mean of the joint angles converges to a tight 
bound around the desired angles. The size of the standard deviation depends on 
the certainty of the prediction of the GP which is influenced by 
the number and the distribution of the
 training points.
\subsection{Experimental Evaluation}
\subsubsection{Setup}
\begin{figure}[tb]
    \centering
    \vspace{0.1cm}
    \includegraphics[width=0.6\columnwidth]{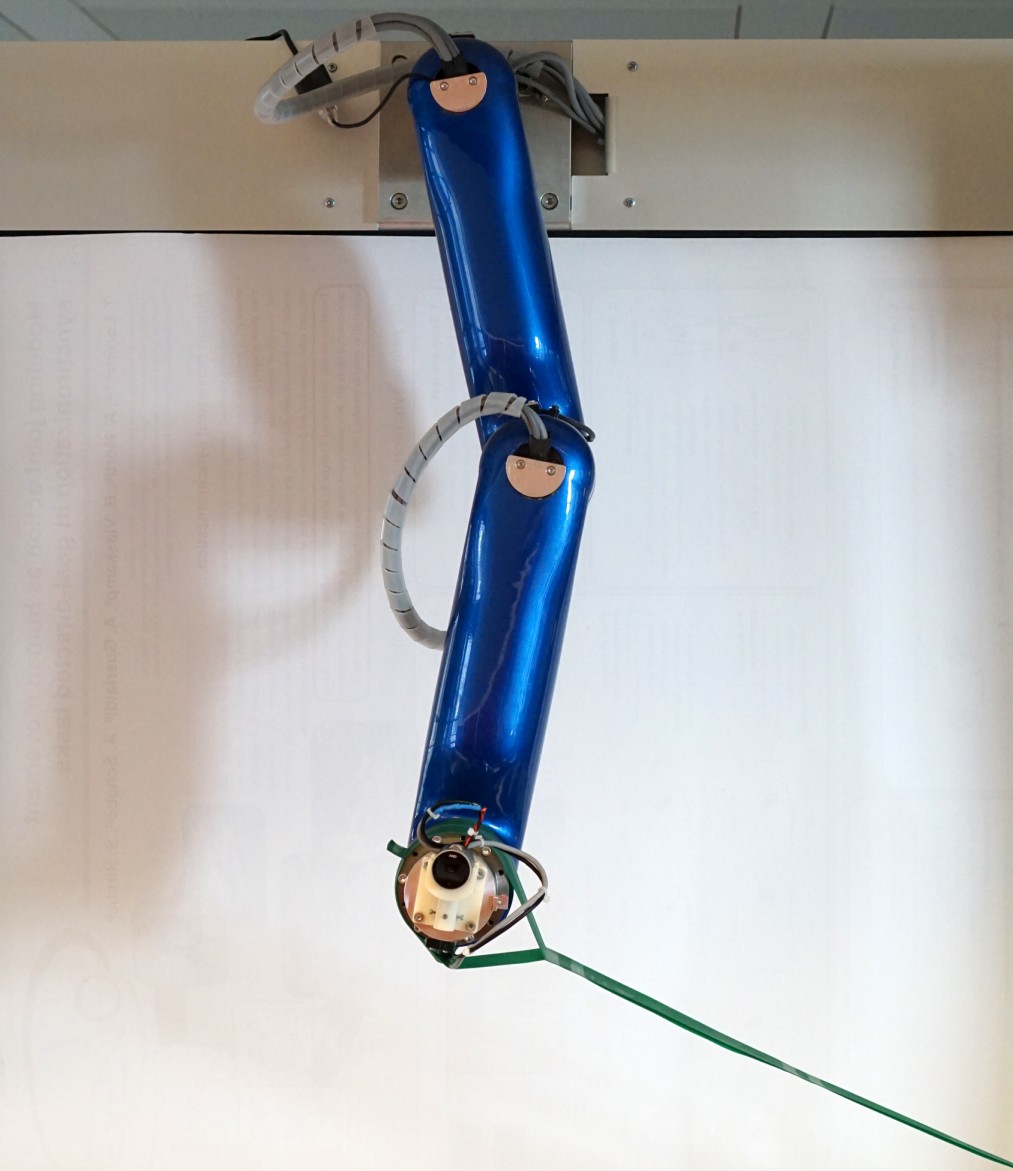}
    \caption{\normalfont A picture of the 3-dof robot CARBO with a rubber 
                  band between the robot's end effector and the ground.} 
      \label{fig:robot}
 \end{figure}

For the experimental evaluation, we use the \mbox{3-dof} SCARA robot CARBO 
as pictured in Fig.~\ref{fig:robot}. The links between the joints have a length of 
$\SI{0.3}{\meter}$. Since the third joint just rotates a camera which is mounted 
as end effector, this joint is fixed for the experiment. A  low level PD-controller 
control enforces the generated torque by regulating the voltage based on a 
measurement of the current (which is approximately proportional to the torque).
The robot manipulates a 
flexible rubber band which is fixed on the right side of the workspace. 
There exists no precise model for the flexible, nonlinear behavior of the rubber 
band, which makes the learning approach necessary. The task is for example
comparable with the handling of rubber seals in the automotive manufacturing.
The desired trajectory follows a sinusoidal shape with a frequency of 
$\SI{1}{\second^{-1}}$ for the first,~$\SI{2}{\second^{-1}}$ for the second 
joint and an amplitude of~$\SI{\pi/5}{}$. The controller is implemented in
MATLAB/Simulink on a Linux real-time system with a sample rate of 
$\SI{1}{\milli\second}$. 
   
\subsubsection{Task evaluation}
For the evaluation of the proposed method, we compare five different 
controllers on the same desired trajectory. 
\begin{itemize}
\item{HG-PD:} A high gain PD controller with the parameters~$K_P^{(HG)} = \text{diag} (800,600)$
             and~$K_D^{(HG)} = \text{diag} (5,5)$ without any feed forward model.
\item{LG-PD:} A low gain PD controller with the parameters~$K_P^{(LG)} = \text{diag} (20,15)$
             and~$K_D^{(LG)} = \text{diag} (5,5)$ without any feed forward model.
\item{CT:} A computed-torque controller based on a friction free model of the 
            robot which is generated from the CAD-model combined with the LG-PD.
\item{CT-SP:} A computed-torque controller based on a friction free model of the 
            robot and a linear model of the rubber band combined with the LG-PD.
\item{CT-GP:} A modified computed-torque controller based on a friction free model of the 
            robot and the trained GP (our approach) in combination 
            with the LG-PD.
\end{itemize}
The high gain approach (HG-PD) is not directly comparable to the other approach 
as it suffers from many disadvantages as discussed in the introduction, but it 
serves as a ''ground truth'' here. It was also employed to generate the 
training data for the CT-GP approach by recording~351 training points corrupted by sensor noise at a 
rate of~$\SI{30}{\milli\second}$ while the robot follows the desired trajectory.
The GPR is implemented with the GPML toolbox \citep{rasmussen2006gaussian}. The 
hyperparameter of the Gaussian
Process are obtained through a gradient-based likelihood maximization.
To obtain the best performance, we employ the deterministic version of our controller, thus using the GP's mean function.
The performance is evaluated using the root mean square error (RMSE) between 
the desired and the real position of the joint angles for all controllers.
\subsubsection{Results}
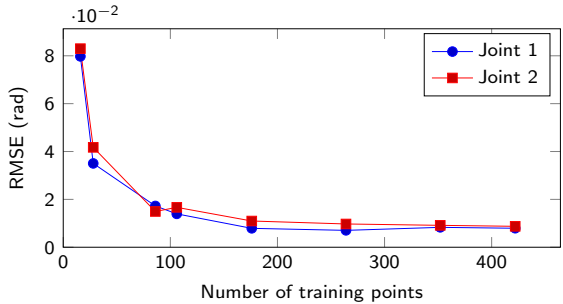
\begin{figure}[tb]
    \centering
    \begin{tikzpicture}[scale=0.9]
\begin{axis}[
   xlabel={Number of training points},
   ylabel={RMSE (rad)},
     width=\columnwidth,
  height=4.8cm,
   xmin = 0, 
   ymin = 0
  ]
\addplot table [x expr=\thisrowno{0}*2, y index=1]{data/figure6_data.dat};
\addplot table [x expr=\thisrowno{0}*2, y index=2]{data/figure6_data.dat};
\legend{Joint 1,Joint 2}
\end{axis}
\end{tikzpicture} 
      \vspace{-0.2cm}\caption{The learning curve of the CT-GP with increasing
                  number of training points.} 
      \label{fig:learning}
\end{figure}

Figure~\ref{fig:error} shows the  RMSE in both joints for the different 
controllers. The low gain controller (LG-PD) performs very poorly, since no
model knowledge is employed. This behavior is improved by adding
 computed torque (CT). 
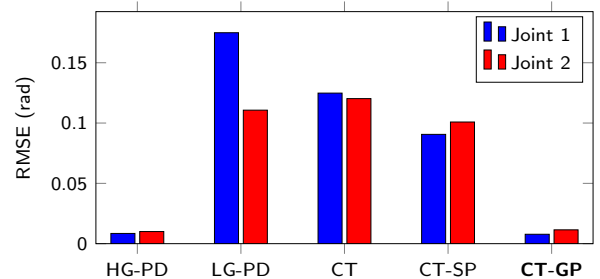
\begin{figure}[b]
    \centering
\vspace{0.2cm}
    \begin{tikzpicture}[scale=0.9]
\begin{axis}[
    ymin = 0,
  ybar,
 xticklabels from table={data/figure5_data_withHG.dat}{name},  xtick=data,
  ylabel={RMSE (rad)},
  ytick={0,0.05,0.1,0.15},
  yticklabels={$0$,$0.05$,$0.1$,$0.15$},
       width=\columnwidth,
  height=5cm,
  ]
\addplot[fill = blue] table [x =xpos, y=value1]{data/figure5_data_withHG.dat};
\addplot[fill = red] table [x=xpos, y =value2]{data/figure5_data_withHG.dat};
\legend{Joint 1,Joint 2}
\end{axis}
\end{tikzpicture} 
      \vspace{-0.2cm}\caption{The RMSE between desired and true joint angles for the 		
      			different control laws. The error of the CT-GP is clearly 
      			smaller than for all other approaches with low gains. The 
      			high-gain approach (LG-PD) has similar RMSE but other undesired 
      			properties and therefore should not be directly compared.} 
    \label{fig:error}
\end{figure} 
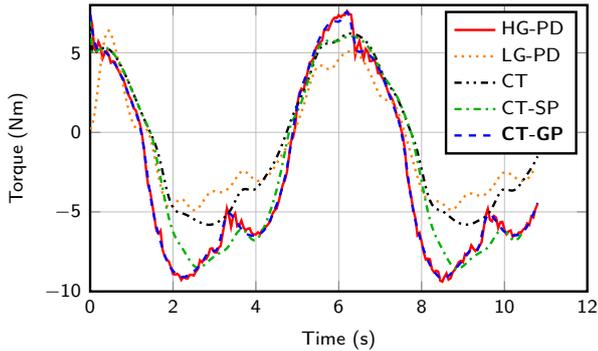
\begin{figure}[tb]
    \centering
    \begin{tikzpicture}[scale=0.9]
\begin{axis}[
  xlabel={Time (s)},
  ylabel={Torque (Nm)},
  line width=1pt,
  grid = major,
  line width=1pt,
       width=\columnwidth,
  height=5.8cm,
  xmin=0, xmax=12, ymin=-10, ymax=8]
\addplot[color=red,no marks] table [x index=0,y index=1]{data/figure3_data_1.dat};
\addplot+[color=orange,style=dotted,no marks] table [x index=0,y index=1]{data/figure3_data_2.dat};
\addplot+[color=black,style=dashdotdotted, no marks] table [x index=0,y index=1]{data/figure3_data_3.dat};
\addplot+[color=green!70!black,style=dashdotted, no marks] table [x index=0,y index=1]{data/figure3_data_4.dat};
\addplot+[color=blue,style=dashed, no marks] table [x index=0,y index=1]{data/figure3_data_5.dat};
\legend{HG-PD,LG-PD,CT,CT-SP,\textbf{CT-GP}}
\end{axis}
\end{tikzpicture} 
      \vspace{-0.2cm}\caption{Applied torque on first joint for different controllers.}
       \label{fig:torque}
 \end{figure} 
 Since the accuracy of the first joint increases a lot, the error of the second joint is slightly 
worse. If the influence of the rubber band is taken into account (CT-SP), the 
accuracy for both joints is improved. Our approach (CT-GP) with low gain
feedback clearly outperforms all other approaches with low gain and is even
competitive with the high gain controller.\\
The applied torque for the first joint is visualized in Fig.~\ref{fig:torque}. The 
CT-GP generates a torque which is very similar to the high gain 
controller, while all others clearly differ. The influence of the amount of 
training data on the performance of our 
approach is shown in Fig.~\ref{fig:learning}. With an increasing number 
of training points, the error is decreasing.
\subsubsection{Discussion}
The simulation and the experiment show that our approach does not only
provide theoretical guarantees, but also shows performance advantages in 
real-world applications. The experiment showed that the feedback gains can be 
reduced by a factor of~$40$ while keeping the performance at a similar level. 
This was not achieved with the best analytically derived physical model for our 
scenario. The simulation showed how highly nonlinear effects (turbulent airflow) 
can also be captured by our nonparametric modeling approach and leads to 
guaranteed diminishing tracking error. 

\section{CONCLUSIONS}
In this paper, we introduce a modified computed-torque control law based on 
Gaussian Process regression (GPR) for robotic manipulators. For 
this purpose, a GP learns the difference between an estimated model 
and the true robot. Afterwards, the control law uses the model and the GPR to 
compensate the robot dynamics. The derived method guarantees stochastic sample path boundedness around zero. If the number of training points tends to infinity, the 
tracking error becomes asymptotically stable.
The proposed control law is of stochastic nature and the convergence occurs in probability in the large. 
Therefore, also its deterministic pendant (the GP's mean function) leads to the stable behavior.


\balance

\bibliography{mybibfile}

\end{document}